\documentclass[twocolumn]{aastex631}
\usepackage{xspace}
\usepackage{graphicx}

\usepackage{amsmath}
\shorttitle{Faraday Rotation and Vacuum Birefringence}
\shortauthors{Krawczynski, Lisalda \& Gammie}
\begin{document}

\input{journals.inp}
\title{The Relative Importance of Faraday Rotation and QED Birefringence 
for the Linear Polarization of X-rays from Mass Accreting Black Holes}

\author{Henric Krawczynski} 
\author{Lindsey Lisalda} 
\affil{Washington University in St. Louis, 
Physics Department, 
McDonnell Center for the Space Sciences, and \newline  
the Center for Quantum Sensors,
1 Brookings Dr., CB 1105, St. Louis, MO 63130} 
\author{Charles Gammie} 
\affil{University of Urbana-Champaign,
Astronomy Department,
202 Astronomy Building, 1002 W. Green Street, 
Physics Department,
235 Loomis Lab, 1110 W. Green Street, Urbana, IL 61801} 

\correspondingauthor{Henric Krawczynski, krawcz@wustl.edu}
\begin{abstract}
The upcoming {\it IXPE} (2-8 keV) and {\it XL-Calibur} (15-75 keV) 
missions will make it possible to measure the linear polarization 
of X-rays from mass accreting stellar mass black holes
with unprecedented sensitivity, enabling the accurate measurement of 
percent-level and in some cases even sub-percent level polarization fractions.
The measurements are expected to constrain the spins, inclinations, and the structure of the accretion flows  
of the observed black holes.
The effects of Faraday rotation and birefringence of the Quantum Electrodynamics (QED) 
vacuum may impact the observable polarization fractions and angles, complicating the
interpretation of the results. 
We estimate the importance of both effects for X-rays from stellar mass and supermassive black holes and 
discuss the implications of the results for the upcoming
{\it IXPE} and {\it XL-Calibur} observations.
\end{abstract}
\section{Introduction}
The X-ray polarimetry missions {\it Imaging X-ray Polarimetry Explorer} (IXPE) 
(2-8 keV, launch in 2021)  \citep{2020SPIE11444E..62S} and {\it XL-Calibur} 
\citep{2021APh...12602529A} (15-75 keV, launches in 2022 and 2023) are expected to
deliver the first high-accuracy measurements of the linear polarization 
fractions and angles of the intermediate and hard X-rays from 
stellar mass black holes. For the brightest extragalactic sources, 
{\it IXPE} may deliver some first polarization detections.
The X-ray polarimetry observations will give us two new observables 
(polarization fraction and angle, or Stokes $Q/I$ and $U/I$, giving the fractional intensities of the polarization 
along vertical/horizontal and diagonal directions, respectively) 
that can be used to test and refine models that were developed based on 
spectral (Stokes $I$) and timing information alone.
Note that the measurement of the circular polarization
(Stokes $V$) of the X-rays from astrophysical sources
is not feasible with the currently available 
technologies.

The X-rays from stellar mass black holes in the thermal state are believed to come from
a geometrically thin, optically thick multi-temperature accretion disk. 
Detailed General Relativistic Magnetohydrodynamic (GRMHD) simulations of such systems
largely validate the results from the early analytical solutions of 
\citet{1973A&A....24..337S} and \cite{1973blho.conf..343N,1974ApJ...191..499P}
in terms of the radial temperature and luminosity profiles \citep{2011ApJ...743..115N,2011MNRAS.414.1183K,2012MNRAS.424.2504Z}.
If geometrically thin, optically thick accretion disks are indeed rather simple, 
then the 2-8\,keV {\it IXPE} observations of the linear polarization fraction 
and angle of the  thermal emission and the 15-75\,keV {\it XL-Calibur} 
constraints on the polarization properties of the power law component
will enable a new way of measuring the inclination 
(angle between the angular momentum vector of the 
inner accretion disk and the observer) \citep{2009ApJ...691..847L}, 
and the spin of the black hole \citep{2009ApJ...701.1175S,2012JPhCS.372a2056D,2012ApJ...754..133K}.
The observations of the intermediate and hard states with {\it IXPE} and
{\it XL-Calibur} will make it possible to constrain the disk and corona 
geometry and location \citep{2010ApJ...712..908S,2013arXiv1301.1957S,2017ApJ...850...14B,2019ApJ...875..148Z}. 
The observations will test the thin disk and corona paradigms, and are important for validating or
refining black hole spin measurement methods \citep[e.g.][]{2013mams.book.....B,2019MmSAI..90..174T,2018GReGr..50..100K}.

Most polarization predictions in the literature assume that the effects 
of Faraday rotation and conversion 
\citep{1986rpa..book.....R,2011MNRAS.416.2574H} and Quantum Electrodynamics (QED) vacuum birefringence 
\citep{1935NW.....23..246E,1936ZPhy...98..714H,Weisskopf:1936,1951PhRv...82..664S,1952PhDT........21T,1971AnPhy..67..599A,1975PhRvD..12.1132T,Heyl.Hernquist:1997,1998ftqp.conf...29D,Hattori.Itakura:2013,Schubert:2000,2015MNRAS.451.3581P,Denisov.etal:2017,Heyl.Caiazzo:2018} on the linear 
polarization properties of the X-rays can be neglected.
This would put X-ray polarimetry in a privileged position as Faraday rotation
(proportional to $\lambda^2$) and Faraday conversion (proportional to $\lambda^3$)
are known to play an important role at longer wavelengths, and 
QED vacuum birefringence (proportional to $1/\lambda$) may 
impact the linear polarization properties at shorter wavelengths \citep{Caiazzo.Heyl:2018}.

\citet{Davis.etal:2009} use detailed shearing box simulations of small segments of stellar mass black holes accretion
disks to estimate the impact of Faraday rotation on the linear polarization
of the thermal X-rays, and find that its effect tapers out around 2 keV.
\citet{Caiazzo.Heyl:2018} raise the possibility that 
QED birefringence affects the observed polarization properties. 
Their calculation assumes mid-disk magnetic field strengths, 
which may exceed those of the regions outside of the accretion disk photosphere 
that the X-rays traverse on their way from their emission to the observer.

In this paper, we present the first estimates of the {\it relative} importance of Faraday rotation and QED birefringence for the polarization of X-rays from mass accreting stellar mass and supermassive black holes.
Our main focus is on these two effects after the emission leaves the photosphere of the accretion disk and traverses the inner region of the accretion flow where magnetic 
fields may be appreciable. 
We start the discussion focused on mass accreting 
stellar mass black holes. We use constraints on the 
Thomson optical depth and the assumption of 
pressure equipartition between the magnetic field 
and the electrons to derive estimates of the 
electron density $n_{\rm e}$ 
and the magnetic field $B$ above and below the accretion disk (Section \ref{s:rev}).
Based on these estimates, we assess the relative importance of 
Faraday rotation and QED birefringence (Section \ref{s:cal}).
The local equipartition between magnetic field and 
electron pressures may not hold in the central regions 
of the accretions flows, and we evaluate the impact 
of much stronger magnetic fields in Section \ref{s:dis}.
We conclude with a discussion of how the situation changes
for Active Galactic Nuclei (AGNs), and the consequences of our findings for the upcoming  {\it IXPE} and {\it XL-Calibur} 
observations.

We use CGS units throughout the paper.
\section{Faraday Rotation and QED Birefringence}
\label{s:rev}
Throughout this section and the next, we assume that the plasma above and below the
inner accretion disk, as well as in the inner accretion disk photosphere is fully ionized. 
For stellar mass black holes and light elements 
(which make up most of the plasma)
this is a good approximation. 
Radiation propagating through a magnetized plasma with electron density $n_{\rm e}$ 
and magnetic field {\bf B} rotates the 
polarization direction by the angle $\Delta \theta_{\rm F}$: 
\begin{equation}
    \Delta \theta_{\rm F}\,=\,
    \frac{2\pi e^3}{m_{\rm e}^2 c^2 \omega^2}
    \int_0^d\,n_{\rm e}\,{\bf B}\cdot\,d{\bf s}
    \label{e:f}
\end{equation}
where $e$ is the electron charge, $m_{\rm e}$ is the mass of the electron,
$c$ is the speed of light, $\omega$ is the radiation's angular frequency, and 
$d$ is the traversed distance, and the result is
positive for a right handed circular polarization 
with the thumb 
pointing  parallel to the wave vector
\citep{1986rpa..book.....R,2011MNRAS.416.2574H}.
We can compare the relative impacts of Faraday rotation and 
electron scattering on affecting the X-ray photons by 
noting that the Thomson optical depth is given by:
\begin{equation}
\tau\,=\,\int_0^d \,n_{\rm e}\,\sigma_{\rm T}\,ds   
\label{e1}
\end{equation}
where $\sigma_{\rm T}$ denotes the Thomson cross section.

Even though the Faraday rotation measure and optical depth are very different 
quantities both of them have a negligible impact on the emerging radiation 
as long as $\theta_{\rm F}\ll1$ and $\tau\ll1$.
The relative importance of both effects depends 
on the magnetic field and the photon frequency $\omega$ 
(or wavelength $\lambda$). 
Assuming {\bf B}$\cdot d${\bf s}\,$\approx B\,ds$, and
and inserting constants we get:
\begin{equation}
    \frac{d\theta_{\rm F}/ds}{d\tau/ds}\,=\,\frac{3}{16\pi^2}\frac{\lambda^2}{e}B.
\end{equation}
Faraday rotation thus becomes significant relative to Thomson
scattering if the ratio exceeds unity which is satisfied for:
\begin{equation}
    B\,>\,\frac{16\pi^2}{3}\frac{e}{\lambda^2}\,\approx\,1.6 \!\times\! 10^6 
    \left(\frac{E_{\gamma}}{{\rm keV}}\right)^2 {\rm\,G\,} 
    \label{z1}
\end{equation}
with $E_{\gamma}$ denoting the photon energy.

The effect of vacuum birefringence depends on the strength of the
magnetic field in units of the critical magnetic field:
\begin{equation}
    B_{\rm c}\,=\,\frac{m_{\rm e}^2\,c^3}{e\hbar}\approx 4.413\times10^{13} {\rm \,G}. 
\end{equation}
For weak magnetic fields with $B\ll B_{\rm c}$, the difference 
of the refractive indices for the radiation polarized 
parallel and perpendicular to the plane of the ambient magnetic 
field {\bf B} and the radiation's wave vector {\bf k} is given by
\citep{1952PhDT........21T,1979ApJ...228L..71C}:
\begin{equation}
n_{||}-n_{\perp}\,=\,\frac{\alpha}{30\pi}\left(\frac{B_{\perp}}{B_{\rm c}}\right)^2
\label{ex}
\end{equation}
where $\alpha\,\approx\,1/137$ is the 
fine structure constant \citep{Zyla:2020zbs}
and $B_{\perp}$ is the strength of the magnetic field {\bf B}
perpendicular to {\bf k}.

\citet{1979ApJ...228L..71C} discuss the implication of Equation~(\ref{ex})
on the polarization of polarized beams for a constant magnetic field, and for magnetic fields 
with varying magnitudes and fixed or variable directions.
The magnitude of the birefringence can be characterized by 
the phase difference incurred by the beams 
with parallel and perpendicular polarization. 
From Equation~(\ref{ex}) we get with $k=\omega n/c$:
\begin{equation}
    \Delta \theta_{\rm QED}\,=\,
    \int_0^d\,(k_{||}-k_{\perp})\,ds\,=\,
    \frac{\alpha\omega}{30 \pi c}
    \int_0^d \left(\frac{B_{\perp}}{B_{\rm c}}\right)^2ds.
    \label{e:q}
\end{equation}
QED birefringence 
is of order unity or larger for:
\begin{equation}
B_{\perp}> 1.6\times 10^8
\left(\frac{d}{2\!\times\!10^7\,{\rm cm}}
\right)^{\,\,-1/2}
\left(\frac{E_{\gamma}}{1\,{\rm keV}}\right)^{-1/2}{\rm G.}
\label{x1}
\end{equation}
Where $d=2\times10^7$~cm is approximately ten times the 
gravitational radius of a 15 solar mass black hole.
The result shows already that very strong magnetic 
fields are needed to render QED birefringence relevant.

Referencing the Stokes parameters such that $Q=+1$ 
corresponds to linear polarization direction in the {\bf B}-{\bf k} plane, 
the effect of vacuum birefringence causes a rotation of {\bf s} 
around the $Q$-axis, so that $U$ rotates into $V$ and vice versa.
When the beam is not entirely monochromatic, and the effect of 
vacuum birefringence is strong ($\Delta \theta_{\rm QED}\gg 1$), the energy dependence of the 
rotation angle leads to a cancellation of the net $U$ and $V$ 
components. 
If the magnetic field direction changes slowly, 
the adiabatic theorem in quantum mechanics implies that 
$Q$ referenced to the slowly rotating {\bf B}-{\bf k} plane 
stays constant, while $U$ and $V$ average to zero
(see \citet{1979ApJ...228L..71C} and also \citet{1978SvAL....4..117G,1979ZhETF..76.1457P,
1979ZhPmR..30..137P,2000MNRAS.311..555H}).

In the following we calculate $\theta_{\rm F}$ assuming
that ${\bf B}||{\bf k}$ holds, and $\theta_{\rm QED}$ for the case of ${\bf B}\perp{\bf k}$.
In actual astrophysical sources, the direction of {\bf B} is expected 
to vary along the photon trajectories, and the
values that we calculate thus overestimate the expected effects.
We will see that $\theta_{\rm F}\ll 1$ and $\theta_{\rm QED}\ll 1$
for good portions of the relevant parameter space.
If one of the values is comparable or much larger than 1, 
the combined effects of the difference in path lengths of the X-rays reaching the observer, 
the integration over a finite range of energies, and the expected variation 
of the magnetic field direction in different portions of the flow are 
likely to lead to a stochastic distribution of the rotation angles of different photons
leading to a strong suppression of the
net polarization.

\section{Estimates of the Impact of Faraday and QED Birefringence 
on Linearly Polarized X-rays}
\label{s:cal}
Using the assumption that the scattering is dominated by a rather high density $n_{\rm e}({\bf x})\approx const$ over
the distance $d$, we can derive a constraint on $n_{\rm e}$
from the measured electron scattering depth $\tau$:
\begin{equation}
    n_{e}(\tau, d)=\frac{\tau}{d \,\sigma_{T}}.
    \label{e2}
\end{equation}
Assuming that the magnetic field pressure $P_B\,=\,B^2/8\pi$ is in 
equipartition with the electron pressure $P_{\rm e}\,=\,n_{\rm e} k_{\rm B} T_{\rm e}$,
we get a magnetic field of 
\begin{equation}
   B(n_{e},T_{e})\,=\,\sqrt{8\pi\, n_{e}\,T_{e}}.
   \label{e3}
\end{equation}
For a given $\tau$, $d$, $T_{\rm e}$, we can thus estimate 
$B$ and $n_{\rm e}$, and use these values to estimate 
$\Delta\theta_{\rm F}$ and $\Delta\theta_{\rm QED}$.
\begin{figure}[t]
\begin{center}
\includegraphics[width=9cm]{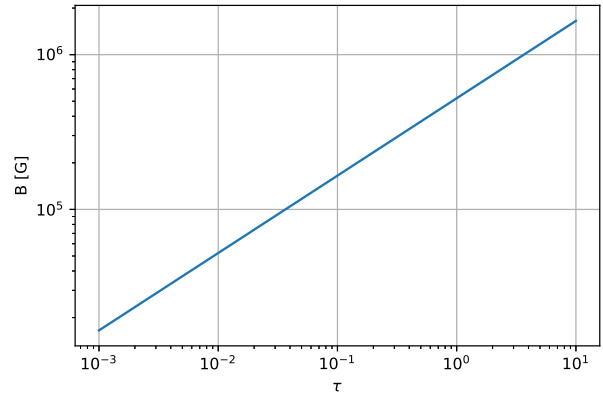}
\end{center}
    \caption{Equipartition magnetic field strength as function of the 
    electron scattering optical depth $\tau$ for photons propagation over 10 gravitational radii of a 15 solar mass black hole, 
    assuming an electron temperature of 100 keV.
    \label{f0}}
\end{figure}

In the following, we present the results for a black hole of mass 
$M\,=$ 15$\,M_{\odot}$ and a gravitational radius 
$r_{\rm g}\,=$ $G\,M/c^2\,=\,$2.2$\times 10^{6}$\,cm (22\,km).
This mass is close to that of the black holes Cyg 
X-1 \citep[$14.8 \pm 1.0 \, M_{\odot},$ ][]{2011ApJ...742...84O} 
and GRS 1915+105 \cite[$12.4+2-1.8\,M_{\odot},$][]{2014ApJ...796....2R}. 

We consider the propagation of X-rays for different optical depths
that the photons traverse before escaping from the inner accretion flow region. 
The photons may come from the
accretion disk or the corona.
Most X-ray photons originate in the central portion of the accretion flow,
and we start with the assumption that the 
$\theta_{\rm F}$ and $\theta_{\rm QED}$ values
build up over a distance $d\,=\,10\,r_{\rm g}$, and that the 
electron temperature is $T\,=$\,100\,keV. 

For $\tau$-values between 0.001 and 10, Equations (\ref{e1}-\ref{e3}) 
lead to magnetic fields between 2$\times10^4$\,G and 2$\times10^6$\,G 
(Figure \ref{f0}). In the remainder of this section, we evaluate 
the consequences of such magnetic fields.  We will discuss the impact of 
stronger magnetic fields in  Section~\ref{s:dis}. 

\begin{figure}[t]
\begin{center}
\includegraphics[width=9cm]{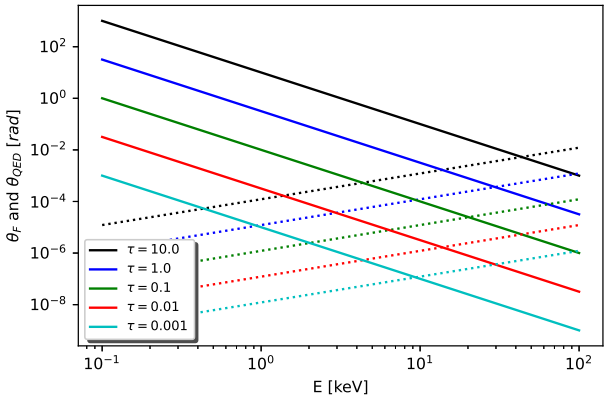}
\end{center}
    \caption{Faraday rotation measure $\theta_{\rm F}$ (solid lines) and
    QED birefringence phase difference $\theta_{\rm QED}$ between the modes 
    parallel and perpendicular to the magnetic field (dotted lines) as a function 
    of the photon energy. 
    The results are shown here for the photon propagation over 10 gravitational 
    radii of a 15 solar mass black hole, assuming an electron temperature of
    100~keV, and a magnetic field in pressure equilibrium with the electrons.
    The color version of this figure is available online. Note for the black and white version that the optical depths decrease from top to bottom. 
    \label{f1}}
\end{figure}

Figure \ref{f1} shows the implied $\theta_{\rm F}$ 
and $\theta_{\rm QED}$ as a function of the photon energy.
Note that our assumption of a constant orientation of {\bf B} along the
path of the photons breaks down for $\tau{\sim}1$ as the photons
change their directions when scattering. The results are still
interesting, as they constrain the relative typical strength 
of Faraday rotation and vacuum birefringence effects.

The figure shows that either the electron density is so low and the magnetic field is so
weak that Faraday rotation and QED birefringence are both negligible, 
or, Faraday rotation is much stronger than QED birefringence for 
all energies below at least 20\,keV.

Varying one parameter at a time, we get the following scaling relations:
$\theta_{\rm F}(d)\propto d^{-1/2}$ and $\theta_{\rm QED}(d)$ is independent of $d$, and
$\theta_{\rm F}(T_{\rm e})\propto \sqrt{T_{\rm e}}$ and $\theta_{\rm QED}(T_{\rm e})\propto T_{\rm e}$.
If the X-rays run through a 100 (10,000) times smaller region of appreciable $n_{\rm e}$ and $B$ with a 
similar net $\tau$ (e.g. within the photosphere of the accretion disk or in a compact wedge
corona above and below the accretion disk), $\theta_{\rm F}$ would be 10 (100) times larger,
and $\theta_{\rm QED}$ would not change. Faraday rotation would thus dominate even 
stronger over QED birefringence. For larger $d$, Faraday rotation becomes less important,
and QED birefringence remains unimportant. Similarly, varying $T_{\rm e}$ from 
temperatures of a few keV all the way to $m_e c^2$ hardly changes the result.

The results above depend on the magnetic field - electron
pressure equipartition assumption to estimate the magnetic field for
a certain optical depth $\tau$. 
Dropping this hypothesis, we can derive the limiting magnetic field
for which the QED birefringence phase difference increases 
more rapidly than the Faraday rotation measure, i.e. for which
$d\theta_{\rm QED}/ds>d\theta_{\rm F}/ds$.
Combining Equations (\ref{e:f}), (\ref{e:q}), and (\ref{e2})  
we get: 
\begin{equation}
B> 1.5\!\times\!10^{9}
\left(\frac{\tau}{0.1}\right)
\left(\frac{d}{2\!\times\!10^7\,{\rm cm}}
\right)^{\,\,-1}
\left(\frac{E_{\gamma}}{1\,{\rm keV}}\right)^{\!-3}{\rm G,}
\label{e5}
\end{equation}
where we used the rough estimates that
{\bf B}$\cdot$d{\bf s}\,$\approx$\,$B\,ds$ and 
$B_{\perp}\approx B$.
This result shows that very strong magnetic fields or
photon energies $>$10\,keV are needed to render 
QED birefringence similarly or more important than Faraday rotation.
\section{Discussion}
\label{s:dis}
Any estimates of the magnetic field strength and electron density
in the tenuous plasma above and below the accretion disk need to be
taken with a grain of salt. The observed energy spectra constrain 
the electron scattering optical depths and densities.
For stellar mass black holes, detailed GRMHD and
shearing box calculations have been carried through.
In the following, we compare our magnetic field 
estimate of $2\times 10^5$\,G derived from the equipartition hypothesis for $\tau=0.1$ (Figure \ref{f1}) 
with various theoretical and observational constraints.

The GRMHD simulations of \citet{2013ApJ...769..156S} of a 10 solar mass black hole 
accreting  at 10\% of the Eddington luminosity indicate magnetic field values 
of between 5$\times 10^5$\,G and 5$\times 10^6$\,G  in the central radiation pressure dominated 
region of the accretion flow along the $\tau=0.1$ contour (see their Figure 3).
That is somewhat higher than the equipartition value derived above, but 
does not change the conclusions of our previous discussion significantly. 
\citet{2009ApJ...691...16H} use 3-D shearing box simulations
to study the vertical structure of the accretion disk of a 
6.62 solar mass black hole accreting at 10\% of the 
Eddington luminosity. In the inner portion of the accretion flow
the radiation pressure and magnetic field pressure are 
in rough equilibrium and exceed the gas pressure by roughly one
order of magnitude. At a radial distance of $r=$30\,$r_{\rm g}$ from the black hole, they
infer a magnetic field strength of $3\times10^6$\,G 
in the scattering photosphere (see their Figures 16 and 18).
Extrapolating their results from 30 to 10\,$r_{\rm g}$ according 
to $B\propto r^{-3/4}$ \citep[][Equation (2.11)]{1973A&A....24..337S}
gives a magnetic field of $8\times10^6$\,G.

For a maximally spinning black hole, the accretion disk extends closer to the black hole, where the magnetic fields are stronger. 

It should be noted,
that none of the existing MHD studies produces the corona self-consistently,
including the Compton scattering and the low Coulomb collisionality of the coronal
plasma, and therefore the vertical temperature and magnetic field structure of
the corona is uncertain. It is likely that magnetic reconnection will convert some of the
magnetic energy of the tenuous plasma into thermal or non-thermal particle energy
\citep[see e.g.][and references therein]{2018ApJ...864...52K}.

We get similar results based on the $\alpha$-disk model of Shakura \& Sunyev.
Based on their Equation (2.11), the mid-disk magnetic field of a 15 solar mass 
black hole is $\leq 6\times10^7$\,G at 10\,$r_{\rm g}$ 
\citep{1973A&A....24..337S}. The result from \citep{2009ApJ...691...16H} 
that the magnetic field outside of the accretion disk photosphere is one order 
of magnitude weaker than in the disk midplane leads again to a magnetic 
field in the mid $10^6$\,G range.

Another magnetic field estimate can be derived from the
observations of an accretion disk wind of the stellar 
mass black hole GRS 1915+105 \citep{2016ApJ...821L...9M}. 
Assuming magnetohydrodynamic or magnetocentrifugal launching 
of the wind, the authors infer magnetic fields of 
between $10^4$\,G and $10^6$\,G at the wind launching 
radius of $\approx$850\,$r_{\rm g}$. 
In the inner, radiation dominated zone of the Shakura \& Sunyaev model, 
the magnetic field scales proportional to
$r^{-p}$ with $p=0.75$ \citep{1973A&A....24..337S}.
Further outwards, the field drops faster reaching an index of
$p=1.31$ in the third zone. For $p=0.75$ the magnetic field 
at $10 r_{\rm g}$ would be about 30 times stronger than at 
850\,$r_{\rm g}$. This route thus gives magnetic field 
values between 10$^5$\,G and several times $10^7$\,G.
Larger $p$-values would lead to even stronger magnetic fields.
Such strong magnetic fields may result from the accretion flow
advecting magnetic fields from the stellar companion and
may play an important role in launching collimated outflows (jets),
see e.g.\,\citep{10.1111/j.1745-3933.2011.01147.x}, and references therein.
\begin{figure}[tb]
\begin{center}
\includegraphics[width=9cm]{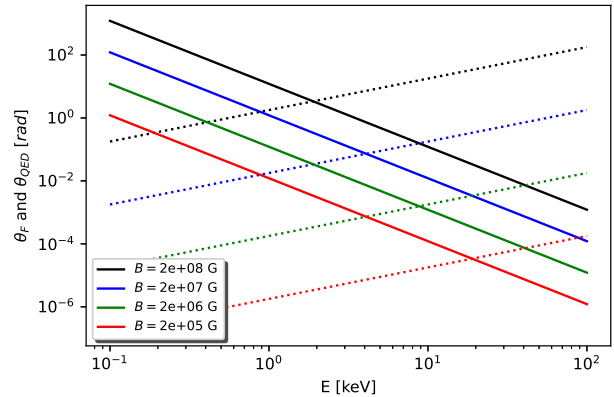}
\end{center}
    \caption{Faraday rotation measure $\theta_{\rm F}$ (solid lines) and QED birefringence phase difference $\theta_{\rm QED}$ 
    for photons traversing a 10 $r_{\rm g}$ large region with
    an electron scattering optical depth $\tau=0.1$ 
    as a function of the magnetic field strength $B$ for
    a 15 solar mass black hole.
    The color version of this figure is available online. Note for the black and white version that the magnetic fields decrease from top to bottom. 
    \label{f2}}
\end{figure}

The discussion above indicates that the X-rays may experience much stronger magnetic fields than those discussed in Section \ref{s:cal}.  
Figure \ref{f2} demonstrates how $\theta_{\rm F}$ and $\theta_{\rm QED}$
computed for $\tau=0.1$ change when we increase the magnetic 
field strength from $2\times 10^5$\,G to values as 
high as $2\times 10^8$\,G. 
For $B\,=$ $2\times 10^7$\,G, Faraday rotation dominates 
up to approximately 4\,keV, at which QED birefringence becomes dominant. 
Note however, that both effects are still rather weak at this energy 
with $\theta_{\rm F}\approx \theta_{\rm QED}\approx4^{\circ}$.
For even stronger magnetic fields of $B\,=$ $2\times 10^8$\,G, 
both effects become important, with QED birefringence starting
to dominate over Faraday rotation at approximately 2\,keV where
$\theta_{\rm F}\approx \theta_{\rm QED}\,\approx\,180^{\circ}$.

Our discussion focused simplistically on the Faraday rotation and
QED birefringence incurred over a $\sim$10\,$r_{\rm g}$ distance close to the black hole. 
We assumed that $n_{\rm e}$ drops quickly as the
photons propagate away from the black hole, rendering the contribution
of large distances to $\theta_{\rm F}$ unimportant. 
Although QED birefringence is independent of $n_{\rm e}$, it
depends quadratically on $B$, so that the net contribution of
large distances to $\theta_{\rm QED}$ may also be negligible.
For regions far way from the central flow, it becomes
increasingly unlikely that the angles between {\bf B} and {\bf k} 
does not change along the photons' trajectories, reducing 
the maximum possible values of $\theta_{\rm F}$ and $\theta_{\rm QED}$.

We can summarize the discussion by noting that Faraday rotation 
and QED birefringence, are rather weak for large regions of the 
likely parameter space of mass accreting stellar mass black holes. 
Extremely strong magnetic fields 
on the order of $10^8$\,G -- which may be required to launch 
jets or winds -- can lead to strong Faraday rotation and QED birefringence effects. For such magnetic fields, both effects could compete for dominance right in {\it IXPE}'s 2\,-8\,keV energy range.
The likely net effect of both effects would be a reduction of the
observed linear polarization fractions. 
Depolarization owing to QED birefringence might be 
responsible for the $<$8.6\% upper limit on the polarization fraction of the 18\,keV-181\,keV emission from the  stellar mass black hole Cyg\,X-1 reported by the {\it POGO+} team \citep{2018NatAs...2..652C}.

How does the situation change for supermassive black holes?
For a $10^9$ solar mass black hole, 10 gravitational 
radii correspond to a distance of $1.48\times10^{15}\frac{M}{10^9 M_{\odot}}$\,cm.
From Equation (\ref{x1}) we infer that 
QED birefringence can play a role for:
\begin{equation}
B_{\perp}> 1.8\times 10^4
\left(\frac{d}{1.5\times10^{15}\,{\rm cm}}
\right)^{\,\,-1/2}
\left(\frac{E_{\gamma}}{1\,{\rm keV}}\right)^{-1/2}{\rm G.}
\label{y1}
\end{equation}
Equation (\ref{e5}) implies that QED birefringence can 
dominate over Faraday rotation for:
\begin{equation}
B>201
\left(
\frac{\tau \times1.5\!\times\!10^{15}\,{\rm cm}}{d}
\right)
\left(\frac{E_{\gamma}}{1\,{\rm keV}}\right)^{\!-3}{\rm G.}
\label{y2}
\end{equation}
Equation (\ref{y1}) sets the more stringent requirement.

The accretion disks of AGNs are able to produce $\sim 10^4$~G
midplane magnetic fields \citep{1986bhmp.book.....T}. 
Again, as for stellar mass black holes, the strength of the 
QED birefringence effect will depend on how
the disk magnetic fields extend into the region 
outside of the disk, the presence of advected magnetic 
fields, and the orientation and uniformity of the magnetic field. 

If Faraday rotation or QED birefringence modify the X-ray 
polarization from stellar mass and supermassive black holes strongly, it will be challenging to disentangle the 
effects of the emission and reflection processes 
from the effects of the propagation of the X-rays through the 
magnetized plasma and QED vacuum.
For observationally well constrained black holes, 
simultaneous broadband spectroscopic observations 
with missions like {\it NICER} and {\it NuSTAR} 
and broadband spectropolarimetric observations with 
missions like {\it IXPE} and {\it XL-Calibur} 
can be used to constrain the geometry of the accretion flow 
based on the spectral information, and the properties of the 
plasma and vacuum through which the X-rays propagate based on the
polarization information
\citep[see also][]{2019BAAS...51c.150K,2019BAAS...51g.181J}.
\section*{Acknowledgements}
The authors thank an anonymous referee for very helpful comments, NASA for the support through 
the grants NNX16AC42G and 80NSSC18K0264, and NSF grants AST 17-16327 and AST 20-07936.

\bibliographystyle{aasjournal}
\bibliography{output}{}
\end{document}